\documentclass[notitlepage,prd,twocolumn,showpacs,preprintnumbers,nofootinbib,superscriptaddress,floatfix,tightenlines]{revtex4-2}
\usepackage{mathrsfs}
\usepackage{amssymb}
\usepackage{amsmath}
\usepackage{graphicx}
\usepackage{amsfonts,amsbsy}
\usepackage{bm}
\usepackage{nicefrac}
\usepackage{xcolor}
\usepackage[breaklinks,colorlinks,urlcolor=blue,citecolor=citcolor,linkcolor=lcolor]{hyperref}

\definecolor{lcolor}{rgb}{0.5,0,0}
\definecolor{citcolor}{rgb}{0,0.3,0.0}
\definecolor{ao(english)}{rgb}{0.0, 0.5, 0.0}
\usepackage{comment}

\newcommand{\p}{{\bf p}}
\newcommand{\pp}{{\bf p'}}

\renewcommand{\k}{{\bf k}}

\def\lra{\leftrightarrow}
\def\onetwo{{1\lra2}}
\def\twotwo{{2\lra2}}
\def \OO {\mathcal{O}}

\def\schemeone {scheme~1}
\def\schemetwo {scheme~2}

\def\schemetwosp {scheme~2 }
\def\md {m}
\def\mg {m_g}

\begin{document}

\title{
Thermalization of non-abelian gauge theories at next-to-leading order}

\preprint{}
\author{Yu Fu}
\affiliation{Key Laboratory of Quark and Lepton Physics (MOE) \& Institute of Particle Physics,Central China Normal University, Wuhan 430079, China}
\author{Jacopo Ghiglieri}
\affiliation{SUBATECH, Universit\'e de Nantes, IMT Atlantique, IN2P3/CNRS,
\\
4 rue Alfred Kastler, La Chantrerie BP 20722, 44307 Nantes, France}
\author{Shahin Iqbal}
\email{smi6nd@virginia.edu}
\affiliation{Key Laboratory of Quark and Lepton Physics (MOE) \& Institute of Particle Physics,Central China Normal University, Wuhan 430079, China}
\affiliation{National Centre for Physics, Quaid-i-Azam University Islamabad, Pakistan}
\author{Aleksi Kurkela}
\affiliation{Faculty of Science and Technology, University of Stavanger, 4036 Stavanger, Norway}

\begin{abstract}
We provide the first next-to-leading-order (NLO) weak-coupling description of the thermalization process of far-from-equilibrium systems in non-abelian gauge theory. 
We study isotropic systems starting from either over- or under-occupied initial conditions and follow their time evolution towards thermal equilibrium by numerically solving the QCD effective kinetic theory at NLO accuracy.
We find that the NLO corrections remain well under control for a wide range of couplings and that the overall effect of NLO corrections is to reduce the time needed to reach thermal equilibrium in the systems considered. 	
\end{abstract}

\maketitle

\section{Introduction}
How do non-abelian gauge fields pushed far from equilibrium  approach  the  thermal  state is a central question  in  several  branches  of  physics. In cosmology, far-from-equilibrium configurations of non-abelian fields may be produced during (p)reheating \cite{Traschen:1990sw,Kofman:1994rk,Kofman:2008zz}, caused by first order transitions \cite{Linde:1978px,Traschen:1990sw,Mazumdar:2018dfl}, and are a necessary ingredient for baryogenesis  \cite{Sakharov:1967dj,DiBari:2021fhs}. In all of these cases, an
understanding of thermalization rates is required for quantitative descriptions of these phenomena \cite{Figueroa:2020rrl,Figueroa:2021yhd}. In the early stages of ultra-relativistic heavy-ion collisions a far-from-equilibrium system of gluons and quarks is created. If and how this system reaches local thermal equilibrium plays a crucial part in the phenomenological modeling of the collisions. The recent discussion about the physical origin of collectivity in smaller collision systems created in p-Pb and light-ion collisions \cite{Citron:2018lsq} further emphasizes the importance of a quantitative understanding of thermalization in far-from equilibrium systems.  Furthermore, connections between systems created in atomic physics experiments and gauge field models are being actively studied (see, e.g., \cite{Banerjee:2012xg, Berges:2014bba,Paulson:2020zjd}).

While first-principles non-perturbative lattice simulation of far-from-equilibrium quantum systems remains elusive, the past years have witnessed progress in methods relying on different approximations --- see \cite{Berges:2020fwq,Schlichting:2019abc} for recent reviews.
On one hand, holographic methods 
have been successful in the description of $\mathcal{N}=4$ Super Yang-Mills theory in the limit of  large number of colors $N_c$ and large t'Hooft coupling $\lambda = g^2 N_c$. These studies have advanced to a mature level, even including sub-leading corrections in the t'Hooft coupling \cite{Grozdanov:2016zjj,Folkestad:2019lam}.
%
On the other hand, weak-coupling methods are available for generic theories and have also been widely studied. The first works studying thermalization of pure Yang-Mills theory from simple initial conditions \cite{Kurkela:2014tea} have been extended to Quantum Chromodynamics (QCD) \cite{Kurkela:2018xxd,Kurkela:2018oqw,Du:2020dvp,Du:2020zqg} and calculations based on this physical picture have been extended to describe systems of enough complexity to be used in realistic 
phenomenological modelling of heavy-ion collisions \cite{Kurkela:2018wud,Kurkela:2018vqr} and even in light-ion collisions \cite{Kurkela:2021ctp}. This picture has also been applied to 
parametric estimates of thermalization times during reheating 
\cite{Davidson:2000er,Harigaya:2013vwa,Mukaida:2015ria}.
These studies have, however, been at best limited to leading order (LO) in the coupling constant and it is important to improve the accuracy --- and in particular, to test the validity and robustness of the weak-coupling expansion --- by finding the first subleading corrections to the weak-coupling results. In this paper we provide the first numerical description of thermalization from simple, isotropic initial conditions at next-to-leading order (NLO). 

A direct diagrammatic description of thermalization is prohibitively difficult due to a need to resum diagrams of all loop orders even to obtain a LO result in $\lambda$ \cite{Jeon:1994if}.
At this order, this resummation can be elegantly performed by considering an effective kinetic theory (EKT) that contains all the necessary processes required for a leading-order description of the evolution of the particle distribution functions $f$
\cite{Jeon:1995zm}. In gauge 
theories, the derivation of the of the collision kernels required for the EKT is further non-perturbative \cite{ Arnold:2002zm}. This arises from the Bose-enhancement of ``soft'' infrared modes at the plasma screening scale $\md^2 \sim \lambda \int d^3p f/p $, whose interactions with the typical ``hard'' particles (with $p \sim \langle p \rangle$) are non-perturbative. This, combined with the well-known soft and collinear divergences of the unresummed QCD cross sections, necessitates a resummation that incorporates the physics of in-medium screening \cite{Braaten:1989mz} and Landau-Pomeranchuk-Migdal (LPM) \cite{Landau:1953gr,Landau:1953um,Migdal:1955nv} suppression in the QCD effective kinetic theory \cite{Arnold:2002zm}. 

The physical picture of EKT can be extended to next-to-leading-order accuracy. The NLO corrections arise from the interactions among the soft modes. The resulting terms are
suppressed only by $\md/\langle p \rangle \gtrsim \lambda^{1/2}$, in contrast to $\lambda$ in vacuum field theory. While various NLO corrections to equilibrium and near-equilibrium quantities have been computed \cite{CaronHuot:2007gq,CaronHuot:2008ni,CaronHuot:2008uw,Ghiglieri:2013gia,Ghiglieri:2018dib,Ghiglieri:2018dgf}, the framework has not until now been pushed to study thermalization of far-from-equilibrium systems. 

In this letter we extend the NLO formulation of EKT to isotropic far-from-equilibrium systems and apply it to numerically describe thermalization of two specific systems initialized with either under- or overoccupied initial conditions studied in LO in \cite{Kurkela:2014tea}.  
In the idealized limit of weak-coupling, thermalization of under-occupied systems (including those created in heavy-ion collisions) proceeds through the process of \emph{bottom-up} thermalization \cite{Baier:2000sb,Kurkela:2011ti}.  The starting point of bottom-up thermalization is an ensemble of too few particles particles $f\ll 1$ with too high momenta $p\gg T$ compared to thermal equilibrium with the final temperature $T$. In the bottom-up process, the collisions among these few hard particles lead to soft radiation that forms a soft thermal bath with a temperature $T_s \ll T$. The further interaction between the hard particles and soft thermal bath eventually causes a radiational break-up of the hard particles that heats the soft thermal bath to its final temperature $T$. We will consider how this picture is quantitatively changed when pushing to finite and small values of $\lambda$. We see that the NLO corrections are under quantitative control for $\lambda \lesssim 10$, and we observe that the NLO corrections make thermalization faster. 

As a second system, we consider an overoccupied, $f\gg 1$ initial state in its \emph{self-similar
scaling solution}, that is, a non-thermal, time-dependent fixed point that is rapidly
reached from any overoccupied initial condition -- see \cite{Berges:2008mr,Berges:2012ev,Schlichting:2012es,Kurkela:2012hp,Kurkela:2011ti,York:2014wja}. 
We find that in this case too NLO corrections bring about a faster
thermalization and that, while a bit larger than for the underoccupied scenario, they remain under control over a wide range of couplings.


\section{setup}
\subsection{Leading Order Kinetic Theory}
In the weak coupling limit $\lambda \rightarrow 0$, the evolution of modes with perturbative occupancies $ \lambda f(p)\ll 1$ and whose momenta are larger than the screening scale $p^2 \gg \md^2$ can be described to leading order in $\lambda f$ by an effective kinetic equation for the color averaged gauge boson distribution function \cite{Arnold:2002zm}
 \begin{equation}
 \partial_t f(p,t) = -\mathcal{C}_{2\leftrightarrow 2}[f](p)-\mathcal{C}_{1\leftrightarrow 2}[f](p).
 \label{locoll}
 \end{equation}

The elastic $2\leftrightarrow 2$ 
scattering and collinear $1\leftrightarrow 2$ splitting parts of the
collision operator --- whose precise forms are given in App.~\ref{sub_lo_kin} --- depend respectively on
 effective matrix elements $|\mathcal{M}|^2$ and  splitting rates $\gamma$ which have been discussed in detail in refs.~\cite{Arnold:2002zm,York:2014wja,Kurkela:2014tea,Arnold:2008zu}. 
The elastic collision term includes LO screening effects by consistently regulating the Coulombic divergence in $t$ and $u$ channels at the scale $\md$. The splitting kernel 
includes the effects of LPM suppression \cite{Landau:1953gr,Landau:1953um,Migdal:1955nv,Migdal:1956tc,Baier:1996kr,Zakharov:1996fv}  which regulate collinear divergences. These effects depend on $\md$
and an effective temperature $T_*$
\begin{equation}
\md^2 = 4\lambda \int_{\bf p }\frac{f_p}{p},
\qquad T_* = \frac{2\lambda}{\md^2}\int_{\mathbf{p}}f_p(1+f_p)
\end{equation}
which are self-consistently calculated during the simulation.
The effective theory contains no free parameters besides the coupling constant $\lambda$.
 Our numerical implementation is the discrete-$p$ method of \cite{York:2014wja}.
\subsection{Next-to-Leading Order Kinetic Theory}
NLO corrections to this kinetic picture have been derived in \cite{Ghiglieri:2015ala}
for a dilute set of high-energy ``jet'' partons interacting with a thermal medium 
and in
\cite{Ghiglieri:2018dib}  at first order in the departure from equilibrium, suited for the determination of transport coefficients. 
These $\mathcal{O}(\sqrt{\lambda})$ corrections arise from the self-interactions of
soft gluons 
with $p\sim \md\sim \sqrt{\lambda}T$ 
 appearing in the internal lines in the diagrammatic computation of the collision kernels. 
At this order,
these soft gluons can be treated as classical fields, retaining only the $T/p$-enhanced part
of their equilibrium distribution, and their contributions can be treated
within the Hard Thermal Loop (HTL) effective theory \cite{Braaten:1989mz}. Furthermore,
they can be treated analytically 
without recurring to brute-force HTL computations,
owing to the light-cone techniques introduced in \cite{Aurenche:2002pd,CaronHuot:2008ni,Ghiglieri:2015ala} (see \cite{Ghiglieri:2015zma}
for a more pedagogical exposition).

These calculations can be extended also to some far-from-equilibrium systems. 
As it is known (see e.g.~\cite{Hong:2010at,Kurkela:2014tea,AbraaoYork:2014hbk,Blaizot:2016iir,Ghiglieri:2018dib}), for $p\ll T_*$, the collinear splittings are 
very effective and rapidly build up a \emph{soft thermal tail}. That is, they ensure that
$f(\md\lesssim p\ll T_*)\approx T_*/p$. This, in turn, implies that, in cases with isotropic
initial conditions, the collision operator can naturally accommodate 
the NLO corrections derived in \cite{Ghiglieri:2015ala,Ghiglieri:2018dib}. The NLO corrections
are suppressed --- with respect to the LO terms in Eq.~\eqref{locoll} --- by a factor of $\lambda T_*/\md$. This arises from the product of the naive suppression factor for loops $\lambda$
with the occupation number at the scale $p\sim \md$, that is $\lambda f(\md) \approx \lambda T_*/\md$\footnote{In equilibrium $\lambda T_*/\md\sim g$ becomes the well-known
suppression factor $g$ of loops at the screening scale $gT$.}. Isotropy
further ensures that the terms which have not been determined in the ``almost NLO''
determination of \cite{Ghiglieri:2018dib} do not contribute here, guaranteeing
that what we are presenting is the full set of NLO modifications.

\begin{figure*}[ht]
    \centering
    \includegraphics[width=0.9\textwidth]{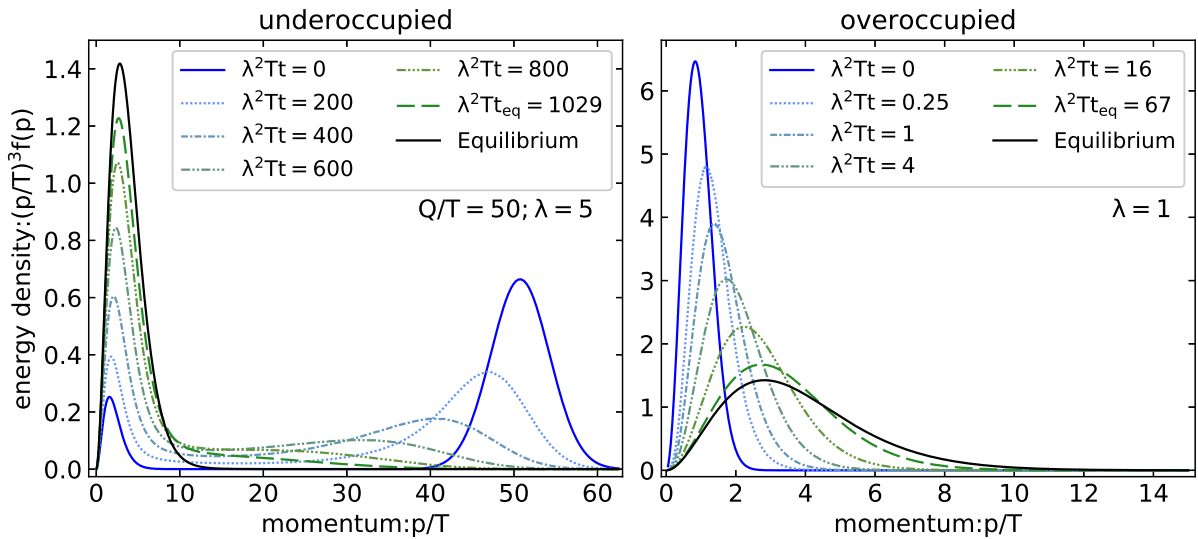}
    \caption{Time evolution from the initial conditions \eqref{eq:initcond}-\eqref{eq:initcondover} 
    in solid blue lines to the final equilibrium state in solid black.
    The dotted and dashed lines show intermediate steps upon solving the
    NLO kinetic theory (\schemetwo). The values of the couplings
    are $\lambda=5$ and $\lambda=1$ respectively.
    \label{fig:timeevo}}
\end{figure*}

These $\OO(\lambda T_*/\md)$ contributions, which we 
discuss in more detail in App.~\ref{sub_nlo_kin}, consist of new scattering
processes and modifications to the LO ones, as  shown in \cite{Ghiglieri:2015ala,Ghiglieri:2018dib}. 
The rate of soft
$\twotwo$ scattering is modified.  This
modification, and an $\OO(\lambda T_*/\md)$ correction to the in-medium 
dispersion, also provide an $\OO(\lambda T_*/\md)$ shift in the $\onetwo$ 
rate. This $\onetwo$ splitting rate must furthermore be corrected wherever
one participant becomes soft 
or when the opening
angle becomes less collinear.  

A rather general 
property of kinetic theory resummations is that it is possible 
to construct collision operators that are equivalent up to 
a given order but differ by subleading corrections. This
was exploited in  \cite{York:2014wja,Kurkela:2014tea}
to construct a LO implementation that is numerically
well-behaved, thanks to a partial resummation of
higher-order effects: a subtraction will thus be needed
to ensure that no double-counting takes place.

We exploit this same property at NLO: as we shall show in detail in
 App~\ref{sub_nlo_kin}, we construct two separate collision
 operators, both including all $\OO(\lambda T_*/\md)$ effects but differing at
 higher orders. We call these two schemes \schemeone\:
 and \schemetwo. The difference in the results obtained from these two,
 as well as their spread from the LO results, can be taken as an 
 estimate of the uncertainty, in particular when extrapolating
 towards regions where the expansion parameters are no longer small.
One such region is thus $\lambda T_*/\md\gtrsim 1$, while another
arises in the region where $p\gg T_*$. 
As is known (see the detailed discussion in \cite{Arnold:2008zu}), the
\emph{formation time} for a 
collinear splitting process 
grows with $p/T_*$, making the splitting process sensitive not just
to the frequent soft scatterings exchanging $q\sim \md$, but also to the
rarer higher-momentum exchanges. For $p/T_*\gtrsim T_*^2/(\lambda \md^2)$
our form of the LO and NLO $\onetwo$ rate, which only includes $q\sim \md$
scatterings, becomes inaccurate. As we elaborate
in App~\ref{sub_nlo_kin}, our first implementation, \schemeone, 
treats these processes with no partial resummation of higher order
effects and the collision kernel is more prone to extrapolate to (unphysical) negative values than
our second, non-strict implementation, \schemetwo.



\subsection{initial conditions}
For the underoccupied initial condition we will use a gaussian form centered around a characteristic momentum scale $Q$, as in \cite{Kurkela:2014tea}. In order to mimic the situation in the last stage of bottom-up thermalization (and for numerical stability), we embed this distribution of hard particles in a soft thermal bath that carries 10\% of the total energy density 
\begin{equation}
 f(p) = A e^{-\frac{(p-Q)^2}{(Q/10)^2}} +   n_B(p, T_{\rm init}),    
 \label{eq:initcond}
\end{equation}
where $A$ and $T_{\rm init}$ are $A \approx (0.419 Q/T)^{-4}$ and $T_{\rm init}/T \approx 0.562$. $n_B$ is the equilibrium Bose--Einstein distribution.

In the overoccupied case we let the system evolve from the scaling solution \cite{York:2014wja}
\begin{equation}
 \tilde f(\tilde p) =\big(0.22 e^{-13.3 \tilde p}+2.0e^{-0.92 \tilde{p}^2}
 \big)/\tilde p,
 \label{eq:initcondover}
\end{equation}
where $ \tilde p\equiv (p/Q)(Q t)^{-1/7}   $ and 
$f(p)\equiv (Q t)^{-4/7}\lambda^{-1} \tilde f(\tilde p)$. For this initial condition
one has $\langle p\rangle \ll T$ and a direct energy cascade from the IR to the 
UV takes place. We choose $Q$ and an initial time $t_0$ such that $f\gg 1$.

\section{Results}
The thermalization processes of systems initialised with Eqs.~(\ref{eq:initcond}) and 
\eqref{eq:initcondover} are displayed in Fig.~\ref{fig:timeevo} for $Q=50$ and $\lambda = 5$ for the underoccupied case (left panel)
and $\lambda=1$ for the overoccupied case (right panel). Both 
are evolved with the \schemetwosp prescription. 

The NLO evolutions of these systems exhibit the same qualitative features as their LO counterparts.
In the case of underoccupied initial conditions, the NLO evolution shows the characteristic features of bottom-up thermalization:
 one can see the hard particles lose energy through the radiational cascade heating the soft thermal bath. Eventually the system thermalizes as the hard particles are quenched in the thermal bath \cite{Kurkela:2011ti}. In the case of the overoccupied initial conditions, the direct energy cascade to the UV seen at LO is also seen at NLO. The
 departure from the scaling solution takes place once $\langle p \rangle \sim T$, corresponding to $f(p)\sim 1$.

In order to determine thermalization times of these systems, we characterise them in terms of 
effective temperatures $T_\alpha$ 
$$
T_\alpha = \left[\frac{2\pi^2}{\Gamma(\alpha+3)\zeta(\alpha+3)} \int \frac{d^3 p}{(2\pi)^3} p^\alpha f(p)\right]^\frac{1}{\alpha + 3}
$$
which all coincide with $T$ in equilibrium but differ for non-equilibrium systems. We then define a (kinetic) thermalization time by demanding that the different effective temperatures are sufficiently close to each other. Specifically, we define the (kinetic) thermalization time using the condition \cite{Kurkela:2018oqw}
\begin{equation} 
\left(T_0(t_{\rm eq})/T_1(t_{\rm eq}) \right)^{\pm4} = 0.9,
\label{eq:condition}
\end{equation} 
where we use "+" and "-" for under- and overoccupied systems, respectively.
For the underoccupied (overoccupied) system in Fig.~\ref{fig:timeevo}, this condition is fulfilled for $\lambda^2 T t \approx 1029$ ($\lambda^2 T t \approx 67$), denoted by the green dashed line. At this point most of the energy is in the thermal bath, rather
than in the initial UV (IR) structure. 

\begin{figure*}[ht]
    \begin{center}
        \includegraphics[width=\textwidth]{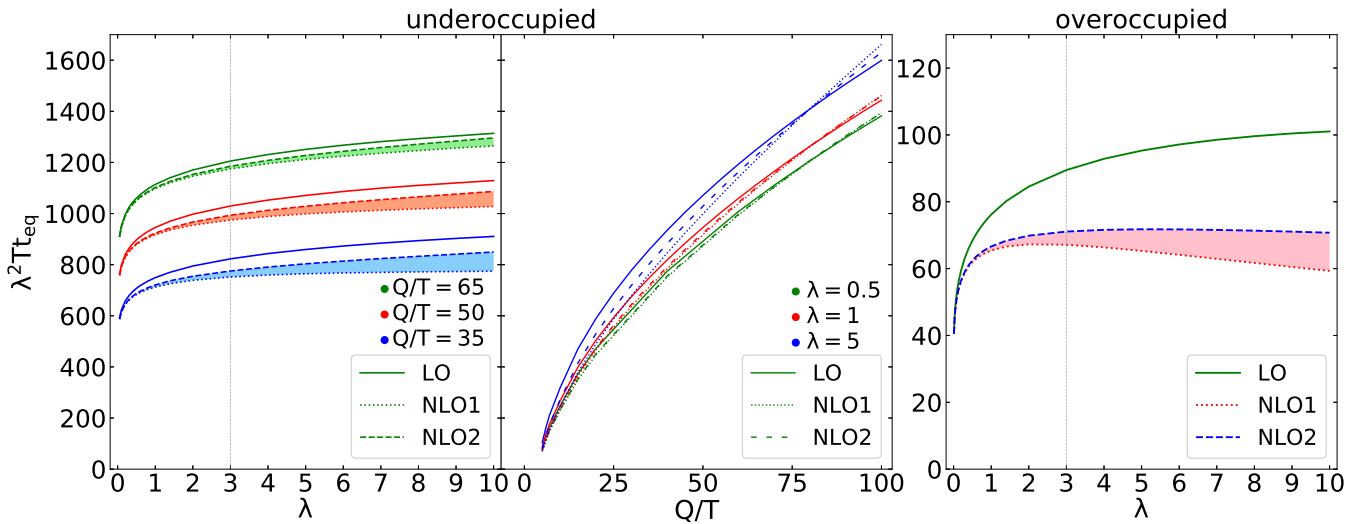}
    \end{center}
   \caption{Equilibration times as a function of the coupling and,
   in the underoccupied case, the initial UV scale. The shaded band between
   the two NLO schemes can be taken as a first indication of the theory
   uncertainty. The coupling $\lambda=3$ for which $m=T$ in thermal equilibrium is indicated by a vertical gray line.
      \label{fig_teq} 
   }
\end{figure*}

We have determined this thermalization time for different values of the coupling constant $\lambda$
and, in the underoccupied case, a variety of initial momenta $Q$, using both the LO as well as the two NLO schemes; the under- and overoccupied-case results are documented in Tab.~\ref{tab:tab1} and \ref{tab:tab2} and displayed in Fig.~\ref{fig_teq}. Our main findings are that 
\begin{itemize}
    \item the qualitative effect of the NLO corrections is to reduce the time required for  thermalization
    \item and that
NLO corrections are well under control for a wide range of coupling constants.
\end{itemize}  
In the regime of small values of $\lambda \lesssim 3$ --- corresponding to $m \lesssim T$ in equilibrium, so that the scale separations assumed in the derivation of the kinetic theory are fulfilled --- the NLO corrections constitute merely a {\bf 5\%} and {\bf 20\%} reduction of the thermalization time in the under- and overoccupied cases. It is reassuring to observe that, in both scenarios, results from the two NLO schemes are close to each other compared to the overall size of the NLO correction. In the $\lambda \rightarrow 0$-limit, the difference between the two NLO schemes vanishes faster than their difference to LO. This  demonstrates that the observed differences from the LO are true NLO corrections and are not contaminated by the scheme differences that affect the result beyond the NLO accuracy.

 Extrapolating to higher values of $3 \lesssim \lambda\lesssim 10$, we see that
 in the underoccupied case the difference between the two NLO schemes becomes 
 comparable to the size the of the NLO correction itself. This
 indicates quantitative sensitivity to corrections beyond NLO. 
 However, taking the difference of the two schemes as an estimate of 
 the uncertainty, we observe that, strikingly, the corrections remain below 10\%-level even for these large value of the coupling. In the overoccupied case the correction reaches 40\%-level,
 with only a moderate spread between the two schemes.

At leading order, the underoccupied thermalization time is parametrically (up to logarithms) of order $t_{\rm eq} \sim (\lambda^2 T )^{-1} (Q/T)^{1/2}$ \cite{Kurkela:2011ti}, related to the democratic splitting time of the particles at the scale $Q$ in a thermal bath with temperature $T$. At NLO, corrections are expected to arise at the relative order $\lambda T/m \sim \sqrt{\lambda}$. We find that that LO thermalization time given in Eq.~\eqref{eq:condition}
is well described for $\lambda < 5$ by a fit%
\footnote{
Note that this thermalization time approximately agrees with that of \cite{Kurkela:2014tea} but differs slightly due to slightly different initial conditions and the precise definition of thermalization time used here. 
}
\begin{equation}
  \lambda^2 T t_{eq}^\mathrm{LO} \approx  (Q/T)^{1/2} (173. + 9.8 \log\lambda)-277.
\end{equation}
For small $\lambda < 1$ and $20 < Q < 80$ the NLO correction in both schemes is approximately given by
\begin{equation}
    \frac{t^\mathrm{LO}_{eq}}{t_{eq} ^\mathrm{NLO}}  \approx 1 + \lambda^{1/2}\left( 0.22-0.05 \log\left(\frac{Q}{T}\right)\right),
\end{equation}
and similarly for the overoccupied case 
\begin{equation}
{\lambda^2 T t_{eq}^\mathrm{LO}} \approx \frac{76.}{1 - 0.19 \log{\lambda}},   \quad
\frac{t^\mathrm{LO}_{eq}}{t_{eq} ^\mathrm{NLO}}  \approx 1 + 0.14 \lambda^{1/2}.
\end{equation}

\section{Conclusions}
The poor convergence of the perturbative series for several different quantities has limited its usefulness in many phenomenological applications. The soft corrections studied here are responsible for this poor convergence for many observables such as transport coefficients \cite{Ghiglieri:2018dgf,Ghiglieri:2018dib} or momentum 
broadening coefficients \cite{CaronHuot:2007gq,CaronHuot:2008ni}. For these quantities
 NLO corrections completely overtake the LO results for $\lambda\approx 10$.
On the contrary, in the present case of isotropic thermalisation, these soft corrections seem to be well under control; the corrections are at most of order 40\% for 
the overoccupied case at $\lambda\approx 10$. These findings are ostensibly in sharp contrast.


However, it is important
to note that \cite{Ghiglieri:2018dib} found NLO corrections to transport
coefficients to be numerically
dominated by the NLO contribution to the isotropization rate governed
by the transverse momentum broadening 
coefficient $\hat{q}$ (which obtains a large positive NLO correction \cite{CaronHuot:2008ni}). 
The key difference with respect to the present case
is that, in an \emph{isotropic} setting, the dependence on $\hat q$ is significantly reduced. 
Instead of explicitly entering the calculation as an isotropisation rate, $\hat q$
only appears in our case as the source of 
$1\leftrightarrow 2$ splittings; it does make their rate larger, but
its numerical effect is moderated by the fact that, parametrically,
the LPM-suppressed $1\leftrightarrow 2$ splitting rate is
$\propto\sqrt{\hat{q}}$, whereas isotropisation is $\propto\hat{q}$.
Furthermore, the other NLO corrections to  splitting arising from a soft
participant, a wider-angle emission or a rarer larger-momentum radiation-inducing 
scattering tend to decrease the
rate, partially cancelling the $\sqrt{\hat{q}}$-driven increase.
This partial cancellation was already seen in the
thermal photon production rate --- another isotropic observable --- which 
also shows moderate NLO corrections \cite{Ghiglieri:2013gia}. 
This is suggestive of 
a pattern which we think deserves further investigations. We note that some of these issues may be ameliorated in thermal equilibrium by non-perturbative determination of the soft contributions developed in \cite{Panero:2013pla,Moore:2019lgw,Moore:2020wvy,Moore:2021jwe}. However, it is currently not known how these methods could be extended to far-from-equilibrium systems.

Lastly, we point out that, when trying to apply our methods to anisotropic systems,
such as one undergoing Bjorken (1D) expansion, we would necessarily
need to include the isotropizing effect of transverse momentum broadening, further
compounded by the emergence of plasma instabilities  \cite{Mrowczynski:1988dz,Mrowczynski:2000ed,Kurkela:2011ti,Kurkela:2011ub,Hauksson:2020wsm,Hauksson:2021okc}. However, in the final stages of the bottom-up thermalization of heavy-ion collisions, the hard particles interact mainly with the isotropic soft thermal bath.  This suggests that the methods developed here may be extended to improve the phenomenological description of the bottom-up hydrodynamization in heavy-ion collision. 




    
\begin{table}
\centering
\begin{tabular}{ccccc||ccccc}
$ Q/T$ & $\lambda$ & $ \hat t^\mathrm{LO}_{eq} $ & $\hat t^\mathrm{NLO1}_{eq} $ & $\hat t^\mathrm{NLO2}_{eq} $ &
$ Q/T$ & $\lambda$ & $\hat t^\mathrm{LO}_{eq} $ & $\hat t^\mathrm{NLO1}_{eq} $ & $\hat t^\mathrm{NLO2}_{eq} $ \\
\hline
20 & 1 & 503.4 & 465.2 & 473.2 &   
35 & 0.1 & 623.4 & 614.5 & 615.7 \\ 
40 & 1 & 818.7 & 784.1 & 791.8 &  
35 & 0.5 & 707.5 & 683.3 & 687.6 \\
60 & 1 & 1060.0 & 1039.1 &1044.4 &
35 & 1 & 749.3 & 712.5 & 720.7  \\
80 & 1 & 1263.9 & 1261.5 & 1263.2 &
35 & 5 & 859.4 & 764.5 & 803.4  \\
100 & 1 & 1443.4 & 1462.2 & 1459.8 & 
35 & 10 & 910.5 & 774.3 & 849.5  \\ 
\hline
20 & 5 & 588.4 & 489.5 & 528.8 &
50 & 0.1 & 798.9 &791.7 & 793.3 \\
40 & 5 & 934.6 & 845.5 & 882.4&
50 & 0.5  & 897.3 & 878.6 & 882.0 \\
60 & 5 & 1193.8 & 1142.4 & 1163.5&
50 & 1 & 945.5 & 916.9 & 923.6 \\
80 & 5 & 1409.5 & 1410.4 & 1408.9&
50 & 5 & 1070.9 & 998.6 & 1028.7 \\
100 & 5 & 1599.2 & 1661.6 & 1630.4&
50 & 10 &1129.1 & 1027.6 & 1086.4\\
\end{tabular}

\caption{Table of thermalization times $\hat t_{eq}  \equiv \lambda^2 Tt_{eq}$ of underoccupied initial conditions with different $Q/T$ and values of the coupling $\lambda$.}
\label{tab:tab1}
\end{table}

\begin{table}[]
    \centering
   \begin{tabular}{c|cccccccccc}
 $\lambda $ &          $ 0.01 $ & $ 0.03 $ & $ 0.06 $ & $ 0.1  $ & $ 0.3  $ & $ 0.6  $ & $ 1    $ & $ 3    $ & $ 6    $ & $ 10 $\\
\hline 
$\hat t^\mathrm{LO}_{eq}  $ & $ 40.9 $ & $ 46.2 $ & $ 50.4 $ & $ 54.0 $ & $ 63.2 $ & $ 70.4 $ & $ 76.2 $ & $ 89.5 $ & $ 97.1 $ & $ 101.0 $ \\ 
$\hat t^\mathrm{NLO1}_{eq}$ & $ 40.5 $ & $ 45.4 $ & $ 49.1 $ & $ 52.0 $ & $ 58.8 $ & $ 62.9 $ & $ 65.4 $ & $ 67.1 $ & $ 64.1 $ & $ 59.3 $ \\ 
$\hat t^\mathrm{NLO2}_{eq}$ & $ 40.5 $ & $ 45.4 $ & $ 49.1 $ & $ 52.1 $ & $ 59.1 $ & $ 63.6 $ & $ 66.6 $ & $ 71.1 $ & $ 71.7 $ & $ 70.7 $\\
\end{tabular}
   \caption{Table of thermalization times $\hat t_{eq}  \equiv \lambda^2 Tt_{eq}$ of overoccupied initial conditions with different values of the coupling $\lambda$.}
\label{tab:tab2}
\end{table}

\begin{acknowledgements}
J.G. acknowledges support by a PULSAR grant from the R\'egion Pays de la Loire. S.I. and Y.F. were supported in part by the National Natural Science Foundation of China under Grant Nos. 11935007, 11221504, 11890714 and 11861131009. We are grateful to Peter Arnold for useful conversations. 
\end{acknowledgements}

\appendix

\section{Definitions and implementations of the kinetic theory}
\label{app_kin_thy}
\subsection{Leading order kinetic theory}
\label{sub_lo_kin}
The precise form of the LO collision operator reads%
\footnote{Our matrix element is related to that of \cite{Arnold:2002zm} by $|\mathcal{M}|^2=\sum_{bcd}|\mathcal{M}^{a b}_{cd}|^2/\nu$, $f = f_a$, and $\gamma = \gamma^{g}_{gg}/\nu$. $\int_{\p}\equiv \int \frac{d^3p}{(2\pi)^3}$ and $\nu=2d_A=2(N_c^2-1)$ for gluons.}
  \begin{align}
& \mathcal{C}_{2\leftrightarrow 2}[f](p)=   \displaystyle \int_{\k,\p',\k'} \hspace{-0.3cm}\frac{|\mathcal{M}(\md)|^2 (2\pi)^4\delta^{(4)}(p+k-p'-k')}{2\;2k \, 2k' \, 2p \,  2p' }\nonumber
\\&\hspace{-3mm}\times \{f_p f_k [1+f_{p'}] [1+f_{k'}] -f_{p'} f_{k'}[1+f_{p}] [1+f_{k}]\},\label{2to2}\\
& \mathcal{C}_{1\leftrightarrow 2}[f](p)  =  \frac{(2\pi)^3}{2p^2} \displaystyle\int_0^\infty dp' dk' \,  \gamma^p_{p',k'}(\md,T_*)  \nonumber \\
 &\times \{f_p [1+f_{p'}] [1+f_{k'}]-f_{p'} f_{k'} [1+f_{p}]\} \delta (p-p'-k')\nonumber \\
 +& \frac{(2\pi)^3}{p^2} \displaystyle\int_0^\infty dp' dk \, \gamma^{p'}_{p,k}(\md,T_*) \,  \delta (p+k-p') \nonumber\\
 &\times
  \{f_p f_k[1+f_{p'}]-f_{p'} [1+f_{p}] [1+f_{k}]\}\label{onetwo}.
 \end{align}
The elastic kernel given in Eq.~\eqref{2to2} depends on the effective in-medium matrix element $|\mathcal{M}(\md)|^2$. As the vacuum elastic scattering has a $1/t^2 \sim \frac{1}{q^4}$ (and $1/u^2$) infrared divergence, with momentum transfer $q =|\p - \pp |$, it makes the soft small angle scattering contribution to the scattering kernel diverge. This divergence is, however, regulated by the the physics of in-medium screening. A prescription that is accurate to leading order was given in \cite{York:2014wja} by the replacement 
\begin{equation}
\label{xidef}
     \frac{(s-u)}{t}\to\frac{(s-u)}{t} \frac{q^2}{q^2 + \xi^2 \md^2},\qquad \xi_\mathrm{LO} = \frac{e^{5/6}}{2\sqrt{2}},
\end{equation}
where at LO  $\xi$ is fixed to $\xi_\mathrm{LO}$, so as to reproduce the LO
longitudinal momentum diffusion coefficient \cite{Ghiglieri:2015ala,Ghiglieri:2015zma}.

The effective medium-induced collinear splitting/merging matrix element $\gamma$ is given by \cite{Arnold:2002ja,Arnold:2002zm}
\begin{equation}
\label{locollinear}
      \gamma^p_{p',k'}(\md,T_*) =\frac{\lambda}{32\pi^4p}\frac{1+x^4+(1-x)^4}{x^3(1-x)^3}
      \mathrm{Im}(\boldsymbol{\nabla}_{\bf b}\cdot {\bm F}({\bf 0})),
\end{equation}
with the momentum fraction $x = k'/p$ and where ${\bm F}({\bm b})$ 
resums an arbitrary number of soft elastic scatterings with the medium.
It depends on two dimensionless variables
\begin{equation}
\label{Meta}
    \hat M \equiv 1 - x + x^2,\qquad    \eta \equiv \frac{p x (1-x) \lambda T_*}{\mg^2},
\end{equation}
where $\mg^2=\md^2/2$ is the LO mass for gluons with $p\gg \md$. 
Parametrically $\eta$ is the ratio squared of the formation time of the splitting process $\tau_{\rm form} \sim \sqrt{E/\hat q}  \sim \sqrt{ \frac{x (1-x) p}{\lambda T_* m^2}}$ and of the elastic scattering rate $\tau_\mathrm{el} \sim 1/\lambda T_* $.
 ${\bm F}({\bm b})$ is the solution to this differential equation \cite{Arnold:2002ja,Arnold:2002zm,Ghiglieri:2015ala}
\begin{equation}
\begin{split}
-2i\nabla_{\bm{b}}\delta^2(\bm{b})&=\frac{i}{2px(1-x)}(\hat{M}\mg^2-\nabla^2_{\bm{b}})\bm{F(b)}\\
&+\frac{1}{2}\Big(C(  b)+C(x b)+C((1-x) b)\Big)\bm{F(b)},
\label{split}
\end{split}
\end{equation}
$C(b)$ is the Fourier transform of the soft scattering rate, 
\begin{equation}
\label{defc}
 C(b) = 
 \int \frac{dq_\perp^2}{(2\pi)^2} (1 - e^{i  {\bm b} \cdot {\bm q}_\perp}) \frac{d  \Gamma(q_\perp)}{d^2 q_\perp}.
 \end{equation}
In an isotropic medium it reads
\begin{align}
\label{clo}
    C(b) = \frac{\lambda T_*}{2\pi}\left( K_0(b \md) + \gamma_E + \log\left(\frac{b \md}{2}\right)\right).
\end{align}
By rescaling ${\bm b}= \tilde{\bm b}/\mg$ and $\bm{F}=2px(1-x)/\mg^2\tilde{\bm{F}}$, the coefficient
of the second line fo Eq.~\eqref{split} becomes proportional to $\eta$. The method
presented in \cite{Ghiglieri:2014kma} is then used for the numerical solution.

\subsection{Next-to-leading order kinetic theory}
\label{sub_nlo_kin}
Let us start by discussing the corrections to Eq.~\eqref{locollinear}. As 
shown in \cite{Ghiglieri:2015ala}, its form remains valid at NLO, but
the LPM resummation in Eq.~\eqref{split} must include two 
$\OO(\lambda T_*/\md)$ corrections. The dispersion relation gets shifted to
$\mg^2{}_\mathrm{NLO}=\mg^2+\delta \mg^2$ and the soft scattering kernel
is modified in  $C_\mathrm{NLO}(b)=C(b)+\delta C(b)$. For an isotropic state
with a $T_*/p$ soft thermal tail, the equilibrium results for $\delta \mg^2$ \cite{CaronHuot:2008uw} and $\delta C(b)$ \cite{CaronHuot:2008ni,Ghiglieri:2013gia} can be used
with the replacement $T\to T_*$, $m_D\to \md$. The former reads
\begin{equation}
\label{deltam}
    \delta \mg^2 = -\frac{\lambda T_*\md}{2\pi }.
\end{equation}
In our first implementation, i.e. \emph{\schemeone}, we treat $\delta \mg^2$
and $\delta C(b)$ as perturbations to their LO counterparts. Hence ${\bm F}$
is perturbed as ${\bm F}_\mathrm{NLO}={\bm F}+\delta{\bm F}$, and the 
latter is computed exactly as in App.~E of \cite{Ghiglieri:2015ala}.\footnote{%
$\bm b$ here corresponds to $p{\bf b}$ there,  ${\bm F}_\mathrm{NLO}$ here corresponds
to $p^3({\bf F}_0+{\bf F}_1)$ there. $\delta C(b)$ can be found in \cite{Ghiglieri:2013gia}.} 
The resulting $\gamma_\mathrm{NLO}=\gamma+\delta\gamma$ can become problematic 
when extrapolated to large values of $\eta$ and $\lambda T_*/\md$. As per its
definition, large values of $\eta$ correspond to formation times larger than the 
mean free time for soft scatterings, so that rarer, harder scatterings, which are not
included in the form~\eqref{clo} of the scattering kernel, would have a chance to
occur. As shown in \cite{Arnold:2008zu}, for $\eta\gtrsim (T_*/\md)^4$
scatterings with $q_\perp\sim T_*$ would need to be included, which is far from trivial
in an off-equilibrium setting. At LO one can however expect, as in equilibrium,
that the approximation introduced by extrapolating Eq.~\eqref{clo} 
to $\eta\gtrsim (T_*/\md)^4$ amounts to an overestimate 
of $\gamma$ at the 10-20\% level. That happens
because large values of $\eta$ privilege the small-$b$ form of $C(b)$, which at 
leading order is approximated by $\lambda T_* \md^2 b^2\ln(1/b\md)$, with a coefficient that varies
in equilibrium by $25\%$ between  $ 1/T\ll b \ll 1/m_D $ and $1/T\gg b$. 

At NLO this translates for large $\eta$ into a strong sensitivity on 
$\delta C(b\ll 1/m)\approx -\lambda^2 T_*^2 b/(32\pi)$, which is 
the Fourier transfor of
 the subleading, $\propto 1/q_\perp^3$, form of the collision 
 kernel for $m\gg q_\perp \gg T_*$. Its negative
coefficient, for large enough $\lambda T_*/m$ and $\eta$, makes $\gamma_\mathrm{NLO}$
negative. We thus propose a second implementation, \emph{\schemetwo}, so that
the difference between the two can be taken as a proxy for the reliability
of these extrapolations. In this second implementation, we do not treat $\delta \mg^2$
and $\delta C(b)$ as perturbations. We rather solve
\begin{align}
\nonumber & -2i\nabla_{\bm{b}}\delta^2(\bm{b})=\frac{i}{2px(1-x)}(\hat{M}\overline{\mg}^2-\nabla^2_{\bm{b}})\bm{\bar{F}(b)}\\
&+\frac{1}{2}\Big(C(x  b)+C( b)+C((1-x) b)\Big)
\left(1+\frac{\delta C}{C}\right)
\bm{\bar{F}(b)},
\label{split2}
\end{align}
where we have defined the mass self-consistently as 
\begin{equation}
    \overline\mg \equiv \sqrt{\mg^2+\frac{\lambda^2T_*^2}{8\pi^2}}-\frac{\lambda T_*}{2\sqrt{2}\pi}\approx \mg\left(1 -\frac{\lambda T_*}{2\pi\md} +\ldots\right),
\end{equation}
i.e. the positive solution to $\overline\mg^2= \mg^2-\lambda T_*\overline\mg/(\sqrt{2}\pi)$, so that, by resumming 
some higher-order terms, it stays positive at large $\lambda T_*/\md$.
In a similar spirit, we have implemented the collision kernel as
\begin{align}
    \frac{\delta C}{C}&\equiv \frac{\delta C(  b)+\delta C(x b)+\delta C((1-x) b)}
    {C(  b)+C(x b)+C((1-x) b)},
\end{align}
so that
$\delta C$ is not treated as a perturbation
in this scheme.
Hence,
the difference between the two schemes, in particular at small to moderate values of $\lambda T_*/m$ and large values of $p/T_*$, is a measure of the uncertainty caused by the lack of harder scatterings 
in the implementation of LPM resummation.

The remaining genuine NLO corrections are 
\begin{enumerate}
\item  wider-angle ``semi-collinear'' $\onetwo$ processes,
\item  contributions to longitudinal momentum diffusion arising from soft legs in $\onetwo$ processes and from soft loops in $\twotwo$ processes.
\end{enumerate}
We implement the two together, following \cite{Ghiglieri:2018dib}.
This amounts to the addition of this extra $\onetwo$ splitting rate
\begin{equation}
\begin{split}
\gamma^{p}_{p'k}\Big|_\mathrm{semi}
=&\frac{\lambda}{64\pi^4 p}\frac{1+x^4+(1-x)^4}{x^3(1-x)^3}\int\frac{d^2h}{(2\pi)^2}\int\frac{d^2q_\perp}{(2\pi)^2}\\
&\delta C(q_\perp,\delta E)
\times\Big[V(1)+V(x)+V(1-x)\Big],
\end{split}
\label{semi}
\end{equation}
where
\begin{equation}
\begin{split}
& \delta E({\bm h}) = \frac{h^2+\hat{M}\mg^2}{2px(1-x)},V(v)=\big(\frac{\bm{h}}{\delta E(\bm{h})}-\frac{\bm{h}+v\bm{q}_\perp}{\delta E(\bm{h}+v\bm{q}_\perp)}\big)^{2},\\
&\delta C(q_\perp,\delta E)=\frac{\lambda T_*\md^2(q_\perp^2+\delta E^2)^{-1}}{(q_\perp^2+\delta E^2+\md^2)}-\frac{\lambda T_*\md^2(q_\perp^2)^{-1}}{(q_\perp^2+\md^2)}
\end{split}
\end{equation} 
In a nutshell,  this implementation subtracts the single-scattering
term of Eq.~\eqref{locollinear} --- the second term in $\delta C(\bm{q},\delta E)$
is precisely $d\Gamma(q_\perp)/d^2q_\perp$ in Eq.~\eqref{defc} --- and replaces
it with a form that keeps track not only of the medium-induced changes
in the transverse momentum of the particles undergoing splitting, but also
of the changes in the small light-cone component of the momentum, i.e. 
$p^0-p^z$ for $p^0\approx p^z\approx p$. Indeed, as shown in
\cite{Ghiglieri:2013gia,Ghiglieri:2015ala,Ghiglieri:2018dib}, for larger emission angles
these changes are no longer negligible with respect to those in transverse
momentum, and give rise to the form shown here. The soft
gluon carries $q^0-q^z=\delta E$ and is no longer 
kinematically constrained to mediating space-like only interactions
with the medium.

Finally, as anticipated in the main text,
we need to avoid double countings. The $\twotwo$ collision kernel in Eq.~\eqref{2to2} integrates
over values of $k,k',p'$ that can be of order $m$, with $q\sim m$ as well. 
In this region the formulation in Eq.~\eqref{2to2}
is no longer accurate. These slices of phase space can be shown to 
be an $\OO(\lambda T_*/m)$ contribution \cite{Ghiglieri:2015ala,Ghiglieri:2018dib},
though obtained with an improper treatment for these soft modes. Thus, this contribution
needs to be subtracted, as it is 
properly included in the NLO contribution to longitudinal momentum diffusion, incorporated
in Eq.~\eqref{semi}. 
This subtraction is analogous to that discussed in App.~B.3 of \cite{Ghiglieri:2018dib}.
Here we perform it by shifting the value of $\xi$ to $\xi_\mathrm{NLO}\approx \xi_\mathrm{LO}+\OO(\lambda T_*/\md)$. We recall that the 
LO value of $\xi$ is fixed by imposing that the expansion of Eq.~\eqref{2to2}
with the replacement~\eqref{xidef} for $\omega\equiv p -p'$ and $q$ much smaller
than $k$ and $p$ matches the LO Hard Loop evaluation of that limit,
which is proportional to the LO longitudinal momentum diffusion coefficient 
\cite{Ghiglieri:2015ala}. To get $\xi_\mathrm{NLO}$ we must now also
expand for $k\sim \omega,q\ll p$, generating a term of relative order $\lambda T_*/\md$.
We then impose that $\xi_\mathrm{NLO}$ cancels this term, yielding
\begin{equation}
    \frac{\lambda \md^2 }{4 \pi p }\ln\frac{\mu}{\mg} =
    \frac{ \lambda \md^2 ( \frac56  + 
    \ln\frac{\mu}{2\sqrt{2} \xi\mg})}{4 \pi  p }+\frac{3 \lambda^2 \md T_* \xi }{(8 \pi)^2  p },
    \label{xinlo}
\end{equation}
where the l.h.s. is what we impose, i.e. the Hard Loop form, with some UV 
cutoff $\mu$, corresponding to the LO longitudinal momentum diffusion term,
while the r.h.s. contains the terms arising from the explicit
expansion of Eq.~\eqref{2to2}. Keeping only the first, leading term 
we recover $\xi_\mathrm{LO}$. We solve Eq.~\eqref{xinlo} self-consistently,
finding $\xi_\mathrm{NLO}$ in terms of the Lambert function $W$ as
\begin{align}
    \xi_\mathrm{NLO}&=-\frac{16 \md \pi}{3 \lambda T_*}W\left(-\frac{3 e^{5/6} \lambda T_*}{
    32 \sqrt{2} \md \pi}\right)\nonumber\\
    &\approx \xi_\mathrm{LO}+
    \frac{3 e^{5/3} T_* \lambda}{128 \pi \md }+\OO\left(\frac{\lambda^2 T_*^2}{\md^2}\right).
\end{align}

\bibliographystyle{apsrev4-1}

\bibliography{main.bib}

\end{document}